# Determining Magnetic Nanoparticle Size Distributions
# from Thermomagnetic Measurements

R.S. DiPietro,[1] H.G. Johnson,[1] S.P. Bennett,[2] T.J. Nummy,[1] L.H. Lewis,[3] and D. Heiman[1]

[1]Department of Physics, Northeastern University, Boston, MA 02115
[2]Department of Mechanical Engineering, Northeastern University, Boston, MA 02115
[3]Department of Chemical Engineering, Northeastern University, Boston, MA 02115

*Abstract*

Thermomagnetic measurements are used to obtain the size distribution and anisotropy of magnetic nanoparticles. An analytical transformation method is described which utilizes temperature-dependent zero-field cooling (ZFC) magnetization data to provide a quantitative measurement of the average diameter and relative abundance of superparamagnetic nanoparticles. Applying this method to self-assembled MnAs nanoparticles in MnAs-GaAs composite films reveals a log-normal size distribution and reduced anisotropy for nanoparticles compared to bulk materials. This analytical technique holds promise for rapid assessment of the size distribution of an ensemble of superparamagnetic nanoparticles.

Magnetic materials are indispensable components in many electronic device applications and are under active investigation to add new functionality for spin electronic devices.[1] One of the key issues for progress in this area is a better understanding of the characteristics of ferromagnets when their size is reduced to nanometer dimensions. While the magnetic properties of an ensemble of nanoparticles can be obtained from magnetization measurements, determination of the particle dimensions and size distribution usually requires careful analysis of electron microscopy images. In view of this limitation, it is attractive to explore new methods to determine nanoparticle dimensions directly from measured magnetic properties. It is known that the average particle size in an ensemble of magnetic nanoparticles can be estimated from magnetic measurements by several methods, such as fitting the field-dependent magnetic moment, m(H), to a classical Langevin function.[2] In some systems, the average particle size of superparamagnetic[3] nanoparticles can be obtained from the thermal blocking temperature, $T_B$, determined from magnetic measurements, which is defined as the temperature that marks the transition from slow magnetic moment relaxation to rapid relaxation within a proscribed measurement time window.[4,5] While these techniques identify the average diameter of an ensemble of magnetic nanoparticles, they do not provide information on the distribution of the sizes of particles.

In this letter we demonstrate a novel method for obtaining the average diameter as well as the distribution of diameters of an ensemble of superparamagnetic nanoparticles using *thermomagnetic* measurements. This transformation method makes use of an approximation to convert the measured temperature-dependent zero-field-cooling (ZFC) magnetic moment, $m_{ZFC}(T)$, into a probability distribution of the particle size. The method was successfully tested on composite structures containing MnAs nanoparticles in a GaAs matrix. Additionally, a comparison of the thermomagnetically-derived size distribution and that obtained from electron microscopy of nanoparticles allowed determination of the uniaxial magnetic anisotropy constant for MnAs. Moreover, it was found that the distribution in particle size conforms to a log-normal distribution function, and that the anisotropy constant is substantially smaller for nanoparticles than for the bulk system.

The method for extracting the size distribution of an ensemble of superparamagnetic nanoparticle from $m_{ZFC}(T)$ data is based on the Nèel model,[3] which describes the relaxation of noninteracting, single-domain magnetic nanoparticles that experience uniaxial anisotropy. Uniaxial anisotropy provides a double-well potential for the two directions of magnetic moment alignment. The wells are separated by an energy barrier $E_B$ that may be overcome by thermal activation with a relaxation time of $\tau = \tau_0 exp(E_B/k_BT)$, where $1/\tau_0$ is the attempt frequency. In the small-field limit $E_B = K_{eff}V$, where $K_{eff}$ is the effective anisotropy constant and $V$ is the particle volume. Since the typical measurement time for a SQUID magnetometer is $\tau \sim 10^2$ s and the attempt time is $\tau_0 \sim 10^{-9}$s, the ratio of anisotropy energy to thermal energy is commonly given by

$$K_{eff}V/k_BT_B=25. \quad\quad\quad (1)$$

Note that changing either $\tau$ or $\tau_0$ by one order of magnitude produces a small change (~10%) in the constant, which rescales $T_B$ or $V$ by that amount, and thus rescales the diameter by only 3%. Within the assumptions of the model, for a measured $T_B$ this equation allows determination of the magnetic volume of particles when the value of $K_{eff}$ is known. Conversely, $K_{eff}$ may be determined if the average particle volume is known. Finally, there are several limitations to this method that must be considered for application: the magnetic field applied for ZFC measurements must be small compared to the anisotropy field so that it does not lower the effective barrier height;[6] it neglects particle-particle dipolar interactions, which increase the apparent blocking temperature;[7] and complications arise from multiaxial (cubic) anisotropy,[3] which reduce the effective barrier height.

The *functional* distribution of the particle volume $f(V)$ or of the particle diameter $f(D)$ can be computed from the $m_{ZFC}(T)$ data. Since each nanoparticle of volume $V$ is characterized by a unique blocking temperature $T_B$ through Eq. (1), the size distribution may be obtained by mapping the distribution in blocking temperature $f(T_B)$ onto $f(V)$. The moment $m_{ZFC}(T)$ of an ensemble of magnetic nanoparticles is given by[8]

$$m_{ZFC}(T) = \frac{H\,m_S^2(T)}{3k_BT}\int_0^T f(T_B)\,dT_B, \quad\quad\quad (2)$$

where $m_S$ is the saturation moment of the ensemble and its temperature dependence can be neglected for materials with Curie temperature $T_C \gg T_B$. The relative distribution function is given by the approximation

$$f(T_B) \propto \frac{1}{T_B^2}\frac{d}{dT}\big[T\,m_{ZFC}(T)\big], \qu\quad\quad (3)$$

or simply

$$f(\mathrm{D}) = f_o \frac{1}{T_B^2} \frac{d}{dT}\big[T\,m_{ZFC}(T)\big], \qquad (4)$$

where $f_o$ is a scaling function.[9,10]

Assuming spherical particles, the distribution in particle diameter $f(\mathrm{D}) = f(T_B)[dT_B/d\mathrm{D}]$ is computed using Eq. (1) with V = $\pi\mathrm{D}^3/6$. The accuracy of the approximation in Eq. (3) was evaluated by assuming log-normal distributions,

$$f(\mathrm{D}) \propto (1/\mathrm{D}\sigma)\,exp\{-[\ell n(\mathrm{D})-\mu]^2/2\sigma^2\}, \qquad (5)$$

using a range of standard deviations σ (relative distribution width). These $f(\mathrm{D})$ were then used to compute $m_{ZFC}(T)$ via Eq. (2), followed by recomputing $f(\mathrm{D})$ again via Eq. (3). It was found that the log-normal $f(\mathrm{D})$ was accurately reproduced in this test. However, the mean particle diameter <D> = $exp(\mu+\sigma^2/2)$ and the distribution width were slightly overestimated, increasing for increasing standard deviation σ. For narrow distributions, σ ≤ 0.3, the overestimation in the mean diameter and distribution width was limited to ≤ 10%, while for broader distributions, σ = 0.4, the overestimations were ~20%.

The transformation in Eqs. (3-4) was applied to composite structures containing MnAs nanoparticles in a GaAs matrix. This material combines a ferromagnet with a semiconductor which makes it attractive for investigating aspects of spin physics.[11,12,13] MnAs holds potential for devices by virtue of its room temperature ferromagnetism and compatibility with GaAs technology. Nanocomposites containing MnAs nanoparticles can be easily fabricated through self-assembly by annealing homogeneous films of $Ga_{1-x}Mn_xAs$.[14,15,16] The annealing parameters can be adjusted to produce varying magnetic properties,[16,17] ranging from small superparamagnetic particles to larger size particles that are single-domain ferromagnets.

To test the thermomagnetic analysis model, composite films containing MnAs nanoparticles were fabricated by annealing homogeneous $Ga_{0.9}Mn_{0.1}As$ films grown at low temperatures by solid-source molecular beam epitaxy (MBE) on GaAs substrates. MnAs nanoparticles were produced by self-assembly after annealing at high temperatures in the MBE chamber. The starting materials in this study were 20 to 50 nm thick films containing 7-10 at% Mn which were grown at 250 ºC. Figure 1 shows the effects of annealing films at different temperatures through measurements of the low-field, temperature-dependent magnet moment, m(T). The starting films were first annealed in air[18] at 250 ºC for 30-60 minutes to maximize the Curie temperature by removing interstitial Mn ions,[19,20] shown by the open circles. This film is ferromagnetic below the Curie temperature of $T_C$=90 K. Samples annealed *in-situ* at temperatures 530-580 ºC in an arsenic flux produced small superparamagnetic nanoparticles, shown by the squares for 530 ºC annealing. The field-cooled m(T) trend decreases paramagnetically as $1/T$ for increasing temperature. Samples annealed at temperatures ≥630 ºC produced larger ferromagnetic MnAs particles with a bulk-like $T_C$ in the range ~340 K, shown by the triangle data markers. These magnetic properties were measured using a superconducting quantum interference device (SQUID) magnetometer (Quantum Design, MPMS XL-5) with the applied field oriented in the plane of the films. The physical size distribution was obtained by manually analyzing scanning electron microscope (SEM) images (see inset of Fig. 2). It is noted that the particles are well-separated in accordance with the assumptions of the model. The structure of the MnAs nanoparticles was not directly determined, however, slow annealing as used here can produce either zincblende or NiAs-type hexagonal structures,[16,17,21] while rapid thermal annealing generally produces hexagonal crystal structures[14,16].

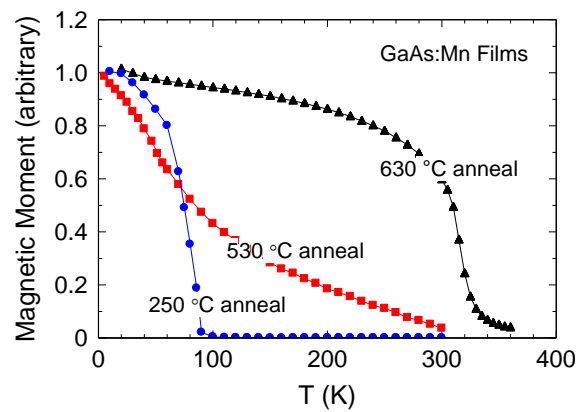

FIG. 1. Magnetic moment of GaAs:Mn versus temperature for 7% Mn. Annealed samples: (*i*) 250 ºC annealed - homogeneous $Ga_{0.93}Mn_{0.07}As$ with $T_c$=90 K (circles); (*ii*) 530 ºC annealed – superparamagnetic MnAs nanoparticles in GaAs matrix (squares); and (*iii*) 630º C annealed - ferromagnetic MnAs nanoparticles in GaAs matrix with $T_c$=330 K (triangles).

Figure 2 shows the measured $m_{ZFC}$(T) data for ensembles of magnetic MnAs nanoparticles acquired by cooling the sample to low temperatures in zero applied field, after which a small field (50-100 Oe) was applied and the magnetic moment measured for increasing temperature. The $m_{ZFC}$(T) data for films annealed at 530 and 580 ºC shows blocking temperature peaks followed by a $1/T$ paramagnetic dependence for increasing temperature. The film annealed at higher temperature has a maximum at a higher temperature, corresponding to a larger average particle size. The distribution of blocking temperatures $f(T_B)$ was computed from the measured $m_{ZFC}$(T) data and Eq. (3). Next, the thermomagnetic distribution of diameters $f(D)$ was computed from $f(T_B)$ and is shown by the solid points in Fig. 3. For comparison, the figure also plots histograms of the physical particle size distributions obtained from SEM images. The thermomagnetic distributions were made to overlap the histograms by selecting a value for the effective uniaxial anisotropy constant.[22] For both films, $K_{eff}$=4x10$^4$ erg/cm$^3$ was used. In Fig 3(a) there is excellent agreement between the two distributions, including the slow decrease on the large diameter side of the peak that is characteristic of a log-normal distribution. The solid curve is a fit to the thermomagnetic data using a log-normal function with a mean nanoparticle diameter of <D>=15.9 nm and standard deviation σ=0.34. In contrast, the film annealed at higher temperature, shown in Fig. 3(b), displays a large difference between the thermomagnetic and physical distributions. While the distribution of physical particle size extends to large diameters, the thermomagnetic distribution is confined to a region of small diameters.[23] This is expected since the larger particles exceed the diameter for superparamagnetism and do not contribute to the ZFC thermal blocking behavior. Thus, superparamagnetism is limited to particle diameters less than 30-40 nm for this system.

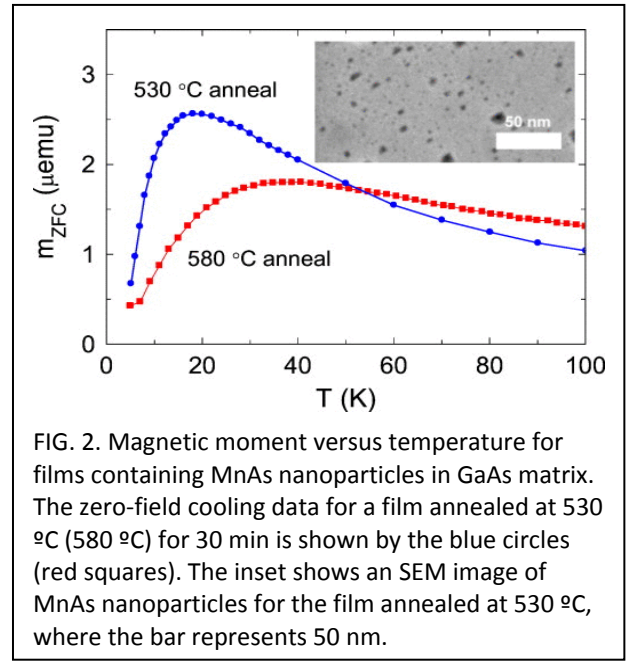

FIG. 2. Magnetic moment versus temperature for films containing MnAs nanoparticles in GaAs matrix. The zero-field cooling data for a film annealed at 530 ºC (580 ºC) for 30 min is shown by the blue circles (red squares). The inset shows an SEM image of MnAs nanoparticles for the film annealed at 530 ºC, where the bar represents 50 nm.

The anisotropy constant is an important materials parameter, and its measurement lends insight into the role of particle-matrix interactions in magnetic nanoparticles, provided that interparticle interactions are minimal. The effective anisotropy constant of MnAs has been reported in the literature for bulk, thin films, and nanocrystals. At room temperature, bulk MnAs is characterized by a value[24] $K_{eff}$=76x10[4] erg/cm[3], while thin films grown on GaAs substrates exhibit anisotropy values[25,26,27,28] ranging from 12x10[4] to 72x10[4] erg/cm[3]. It was also found that for MnAs nanocrystals somewhat smaller than those of this study, the anisotropy values[29] were in the range 14x10[4] to 18x10[4] erg/cm[3]. The smaller anisotropy for embedded MnAs nanocrystals could arise from a modified crystal structure, lattice mismatch at the nanoparticle surface, matrix-induced strain, or electric field effects at the particle-matrix interface. Additional studies are required to examine this aspect.

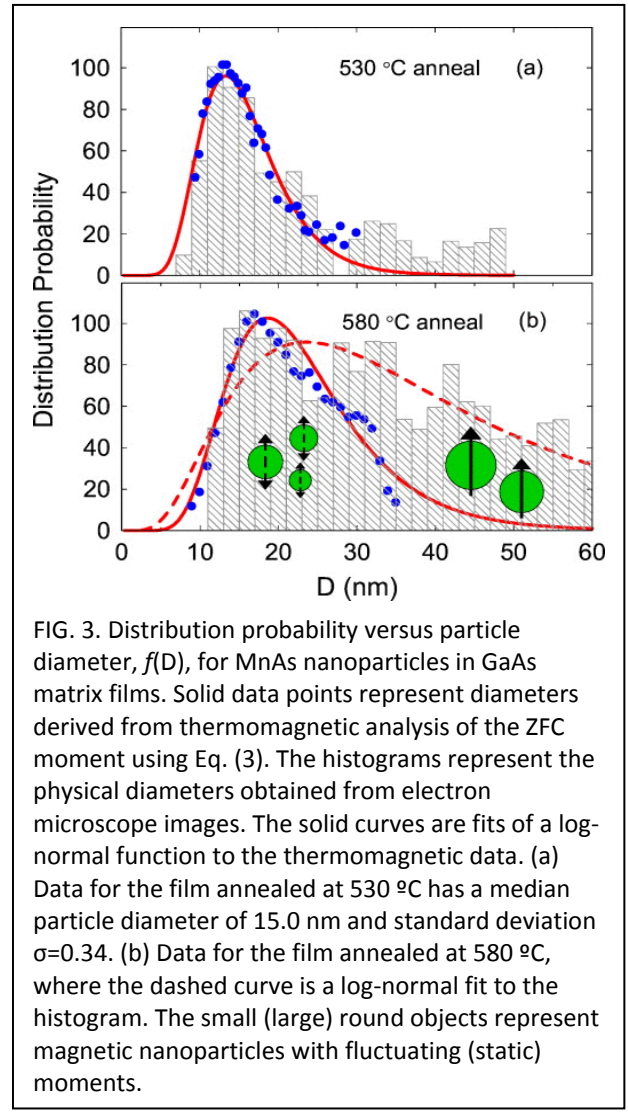

FIG. 3. Distribution probability versus particle diameter, $f$(D), for MnAs nanoparticles in GaAs matrix films. Solid data points represent diameters derived from thermomagnetic analysis of the ZFC moment using Eq. (3). The histograms represent the physical diameters obtained from electron microscope images. The solid curves are fits of a log-normal function to the thermomagnetic data. (a) Data for the film annealed at 530 ºC has a median particle diameter of 15.0 nm and standard deviation σ=0.34. (b) Data for the film annealed at 580 ºC, where the dashed curve is a log-normal fit to the histogram. The small (large) round objects represent magnetic nanoparticles with fluctuating (static) moments.

In conclusion, it is shown that thermomagnetic data obtained from an ensemble of superparamagnetic nanoparticles can be transformed into a distribution in particle size. The transformation relies on ZFC magnetic data that is readily obtained during characterization of the magnetic properties of the nanoparticles. In addition, the nanocrystal anisotropy can be investigated by combining the thermomagnetic distribution with the physical size distribution. The application of this method to MnAs nanoparticles indicates that the anisotropy constant can be significantly modified for nanometer size particles.

ACKNOWLEDGMENTS
This work supported by the National Science Foundation grant DMR-0907007. We thank W. Fowle for assistance with the electron microscopy studies.